\documentclass{andromedaone}       

\journal{NDM}
\vol{2020}
\jyear{Egypt}
\pages{Hurgada} 

\received{xx January 2020}
\published{xx April 2020}

\def\be{\begin{equation}}
\def\ee{\end{equation}}
\def\bea{\begin{eqnarray}}
\def\eea{\end{eqnarray}}
\def\met{\not{\!{\rm E}}_T}
\def\zp{Z^\prime}

\usepackage{graphicx}

\begin{document}

\title{Prospect of the Electroweak Scale Right-handed neutrino model in the Lifetime Frontier}

\author{Shreyashi Chakdar,\auno{1} P.Q Hung,\auno{2}}
\address{$^1$ Department of Physics, College of the Holy Cross, 1 College Street,Worcester, MA 01610, USA}
\address{$^2$Department of Physics, University of Virginia, Charlottesville, VA 22904-4714, USA}

\begin{abstract}
Motivated by the null results of the BSM searches in the post-Higgs era of the LHC, our current approach is to look for new physics shifting from theory driven search strategies to signature driven ones. One possible direction might come from investigating the long-lived particles (LLP’s) present in various theoretical scenarios through the newly formed “Lifetime frontier”. In this talk, I discuss a non-sterile right-handed neutrino model consisting of EW-scale Majorana masses, having signals with large displaced vertices arising in both the fermion and scalar sectors.The characteristic features in this model, the displaced vertices, i.e. several charged tracks originating from a position separated from the proton interaction point has to be greater than a mm and can be as long as order of centimeters. These events originating from the decays of the mirror fermions produce promising signatures at the LHC environment due to the low associated backgrounds. We discuss the experimental implications and possible search strategies in this framework and LHC’s potential to unravel these underlying events.
\end{abstract}

\maketitle


\section{Introduction}

The absence of any new discoveries at the LHC in the post 125 GeV Higgs era have led both the theory and experimental particle physics community to ponder on the question of where and how to effectively look for anticipated new physics signals. As it is apparent that LHC is going to take orders of magnitude of more data in the next 20 years to come with significant more luminosity and there are bigger and powerful linear and hadron accelerators to be built near the horizon, the question we are all trying to answer is if we are looking for hints of the Nature at the right places, using right tools. One of the new directions might come through the newly acquired interest in the Long lived Particles opening up a new "Lifetime Frontier" in LHC searches. In this talk, I will look at a framework that utilizes the LLP sector pointing to a new direction to look for the new physics and connects the neutrino, dark matter and collider physics. \\
So far, most LHC searches for new physics are based on search strategies that look for prompt objects, primarily concentrated on regions of the detector which are close to the collision point in the beam pipe at distances mostly $< O(1mm)$. Perhaps the new physics is hidden somewhere in regions of the detector which are not included in the present search algorithms, e.g. at distances greater than a few millimeters at displaced vertices? Now, from the neutrino oscillation data, the fact that neutrinos have tiny mass is now indisputably proven and although its nature is still unknown, this is the only evidence so far proving the existence of physics beyond the SM (BSM). The big question we are interested in this talk is about if and how the nature of the neutrinos be detected at the LHC through this LLP searches or to go back one step, can neutrino signals even be seen in the Electroweak energy scales accessible to the current colliders? \\
Among many proposals available in the market for the existence of tiny neutrino masses, the most popular one is the so-called seesaw mechanism. In this seesaw mechanism, we postulate the existence of right-handed neutrinos $\nu_R$ involving two mass terms: the Dirac mass term $m_D\bar{\nu_L}\nu_R + H.c$., and the Majorana mass term $M_R\nu_R^T\nu_R$. It is to be noted that the aforementioned mass terms involve a mixing between left-handed and right-handed neutrinos. If one assumes $m_D << M_R$, the diagonalization of a 2x2 matrix yields the eigenvalues: $m_\nu = ({m_\nu^{D}})^2/M_R \approx O (< eV )$ and $M_R = ?$. Some questions we are interested in are about the physics mechanisms behind the generation of $m_D$ and $M_R$, how massive are $\nu_R$’s and do they interact with the Gauge (W and Z) bosons of the SM.
In grand unified models using the "seesaw scenarios" where the SM is embedded in larger gauge groups (SO(10) for example), "sterile" $\nu_R$’s are SM singlets and $M_R$ is naturally of $O(\Lambda_{GUT} \approx 10^{16} GeV$) making $\nu_R$’s inaccessible as they do not couple to SM gauge bosons. Model building tricks can lower the scale of $M_R$ significantly but not enough for $\nu_R$’s to be produced at current high energy accelerators. On the other hand, in the left-right symmetric model framework ($SU(3)_C \times SU(2)_L \times SU(2)_R \times U(1)_{B-L}$) where $\nu_R$’s are $SU(2)_R$ doublets but $SU(2)_L$ singlets, and thus has more potential for the discovery of $\nu_R$’s. However, this scenario is depended to the gauge bosons of $SU(2)_R$, $W_R$, no signs of which has been seen at the LHC so far.\\
In this talk, I discuss a framework, in which $\nu_R$’s are non-sterile and interact with W and Z with the same strength as that of other known SM quarks and leptons and $M_R$ is proportional to $\Lambda_{EW}$. It is to be noted that there are no physical principles forbidding or testifying for a non-sterile $\nu_R^\prime$ and most significantly this could be a testable or falsify-able scenario in the current collider energy scales. This $EW\nu_R$ framework \cite{Hung:2006ap} contains distinguished, low background LLP signatures that could point to a new direction of NP detection.

\section{{EW$\nu_R$ Framework relevant for LLP signals}}
The EW-scale $\nu_R$ model is basically the SM supplemented with extended fermionic and scalar sectors: For every SM left-handed doublet, there is a mirror right-handed doublet, and for every SM right-handed singlet, there is a mirror left-handed singlet. The gauge group is $SU(3)_C \times SU(2)_W \times U(1)_Y$; there is no subscript "L" in $SU(2)$ as the EW-scale $\nu_R$ model accommodate both SM fermions and the mirror counterparts of opposite chirality. The energy scale characterizing the EW-scale $\nu_R$ model is still the electroweak scale $\Lambda_{EW} \sim 246 GeV$. This is the reason for the Majorana mass of the right-handed neutrinos to be bounded from above by the electroweak scale and, as a consequence, for its accessibility at colliders such as the LHC. \\
1. \textbf {The fermionic sector} constitutes of 

SM: $l_L = \left(
	  \begin{array}{c}
	   \nu_L \\
	   e_L \\
	  \end{array}
	 \right)$; Mirror: $l_R^M = \left(
	  \begin{array}{c}
	   \nu_R^M \\
	   e_R^M \\
	  \end{array}
	 \right)$. 
	 
SM: $q_L = \left(
	  	 \begin{array}{c}
	   	  u_L \\
	     	  d_L \\
	  	\end{array}
	 	\right)$; Mirror: $q_R^M = \left(
	  	 \begin{array}{c}
	   	  u_R^M \\
	     	  d_R^M \\
	  	\end{array}
	 	\right)$.

along-with the singlets: SM: $e_R; \ u_R, \ d_R$; Mirror: $e_L^M; \ u_L^M, \ d_L^M.$\\
\noindent 2. The \textbf {scalar sector} in this framework consists of singlets, doublets and triplets higgses.\\ 
a) The singlet scalar Higgs $\phi_S$ with $\langle \phi_S \rangle = v_S$: $\phi_S$ with  $v_S < v_M$ (triplet vev) is primarily important as this singlet scalar connects the SM and Mirror worlds and is a contender for a light keV scale dark matter candidate under study.  It also plays a crucial role in the LHC search for mirror fermions through the displaced vertices (LLP). 
It is important to note that from an updated analysis of MEG $\mu \rightarrow e \gamma$ constrains $g_{Sl} < 10^{-4}$ (typically) which gives $v_S \sim O(GeV)$ \cite{Hung:2015nva}. As a result, the Dirac mass appearing in the seesaw formula previously referred, namely $m_\nu^D = g_{Sl} v_S $, has to be less than $10^{-4}$ GeV in order for $m_{\nu} < O(eV)$ because $M_R \sim O(\Lambda_{EW})$. Nevertheless, one can easily obtain the mass of the physical Higgs singlet scalar mass to be smaller than GeV scale and has been thus ignored in the  phenomenological analysis discussed next.\\
b) Now, let us come to the doublet Higgs sector which consists of: $\Phi_1^{SM} (Y/2 = -1/2)$ with vev  $<v_1>$ and $\Phi_2^{SM}(Y/2 = 1/2)$ with vev $<v_2>$ only coupling to the SM fermions (respectively to the down and up sectors), while in the mirror doublet sector consists of another two doublets $\Phi_{1M} (Y/2 = -1/2)$ with vev $<v_{1M}>$ and $\Phi_{2M} (Y/2 = 1/2)$ with vev $<v_{2M}>$ that couples only to corresponding mirror fermions.  \\
c) Higgs triplets sector consists of a complex triplet $\widetilde{\chi} \ (Y/2 = 1)  = \frac{1}{\sqrt{2}} \ \vec{\tau} . \vec{\chi} = 
	  \left(
	  \begin{array}{cc}
	    \frac{1}{\sqrt{2}} \chi^+ & \chi^{++} \\
	    \chi^0 & - \frac{1}{\sqrt{2}} \chi^+\\
	   \end{array}
		  \right)$ with $\langle \chi^0 \rangle = v_M$.
and a real triplet $\xi \ (Y/2 = 0)$ needed in order to restore Custodial Symmetry with $\langle \xi^0 \rangle = v_M$ (this guarantees $M_W^2 = M_Z^2 cos^2{\theta_W}$ at tree level)
Here all the VEVs statisfy: $v_{1}^2+v_{1M}^2 + v_{2}^2+v_{2M}^2 + 8 v_{M}^2= v^2 \approx (246 Gev)^2$. For detailed treatment see (\cite{Hoang:2014pda})\\		
3. \textbf {Relevant interactions for LLPs}
i) Majorana Neutrino Mass 
$L_M = g_M l^{M,T}_R \sigma_2 \tau_2 \tilde{\chi} l^M_R$, from this we obtain the Majorana mass $ M_R = g_M v_M$ with $M_Z/2 < M_R < O(\Lambda_{EW} \simeq 246GeV$). Energetically accessible at the Current LHC. 
ii) Dirac Neutrino Mass: The singlet scalar field $\phi_S$ couples to fermion bilinear
$L_S = - g_{Sl} \bar{l}_L \phi_S l_R^M + h.c. $ giving Dirac neutrino mass $m_\nu^D = g_{Sl} v_S $.This is an interaction that is crucial in the decay of a mirror quark/lepton into a SM quark/lepton in association to the singlet scalar field $\phi_S$ whose accessibility as the LLP signals depend on the smallness of Yukawa couplings $g_{Sq}/g_{Sl}$ (bounded by $g_{Sq}< g_{Sl}< 10^{-4}$).

\section  {Search for Long lived mirror fermions: Possible signatures}
\subsection{{Mirror Quark signatures}}
In this section, we are interested in a condensed discussion of the signatures of the mirror fermions, in particular the lightest mirror quark (lepton), the decay mode that is allowed is $q^M\to q \phi_S~{\rm or}~b \phi_S$. As mentioned before, the singlet scalars are assumed to be much lighter than the quarks (both SM and mirror) and we will neglect their masses in the computation of the decay width. The expression for the decay width is as follows, \cite{Chakdar:2015sra} 
\begin{center} $\Gamma(q^M\rightarrow q+\phi_S) = (g_{Sq}^2 m_{q^M}/{64\pi}) (1- {m_q^2}/m_{q^M}^{2})(1+ m_q/{m_{q^M}} - {m_q^2}/{2m_{q^M}^2}) $
\end{center}
The decay length is dependent on the Yukawa $g_{Sq}$, in which one finds the Yukawa coupling and various mixing angles. Since the decay length is $\gamma \, \beta \, \hbar \,c/\Gamma(q^M\rightarrow q+\phi^{\star})$, one easily imagine that it can be {\em macroscopic} i.e. $> 1\, mm$ if $g_{Sq}$ is sufficiently small; such macroscopic decay length can presently be missed due to the nature of the prompt decay algorithms of CMS and ATLAS.

A typical decay chain of a mirror quark starts with the heaviest mirror decaying into a lighter one plus a W, e.g. $u_i^M \rightarrow d_j^M + W$. The lightest mirror quark can only decay into a SM quark along-with the singlet scalar $\phi_S$. Now, since the decay length of a ”free” lightest mirror quark ($> 1mm$) is typically much larger than a hadronic length ($\simeq1fermi$), the gluon fusion process first gives rise to mirror meson ($\bar{q}q$) which subsequently decays, at a displaced vertex, into a pair of SM quark and antiquark. Since mirror quarks are heavy, the formation of a mirror meson \cite{Duong} can be accomplished, to a good approximation, by a QCD Coulomb-like potential $V (r) \simeq -4\alpha_s(r)/3r$.  It is straightforward to compute the mirror meson wave function at the origin which allows us to obtain the production cross section and the decay rate. Basically, the relevant process is $\sigma(gg \rightarrow \eta^M \rightarrow q\bar{q}$) where $\eta^M$ stands for ”mirror meson”.  In $\eta^M \rightarrow q\bar{q}$, there is no missing energy in the form of $\phi_S$ because $q$ and $\bar{q}$ exchange $\phi_S$ and transform into q and $\bar{q}$. From Fig. 1, one notices that mirror mesons decay well inside a typical silicon vertex detector and beyond for a range of mass ($m_\eta M \simeq 2m_qM$) and for $g_{Sq} < 10^{-4}$. The lightest mirror mesons are long-lived! Other kinds of mesons and even ”baryons” involving one or more mirror quarks are under study. \begin{figure}[hbt!]
\centering
 \includegraphics[width=0.4\textwidth]{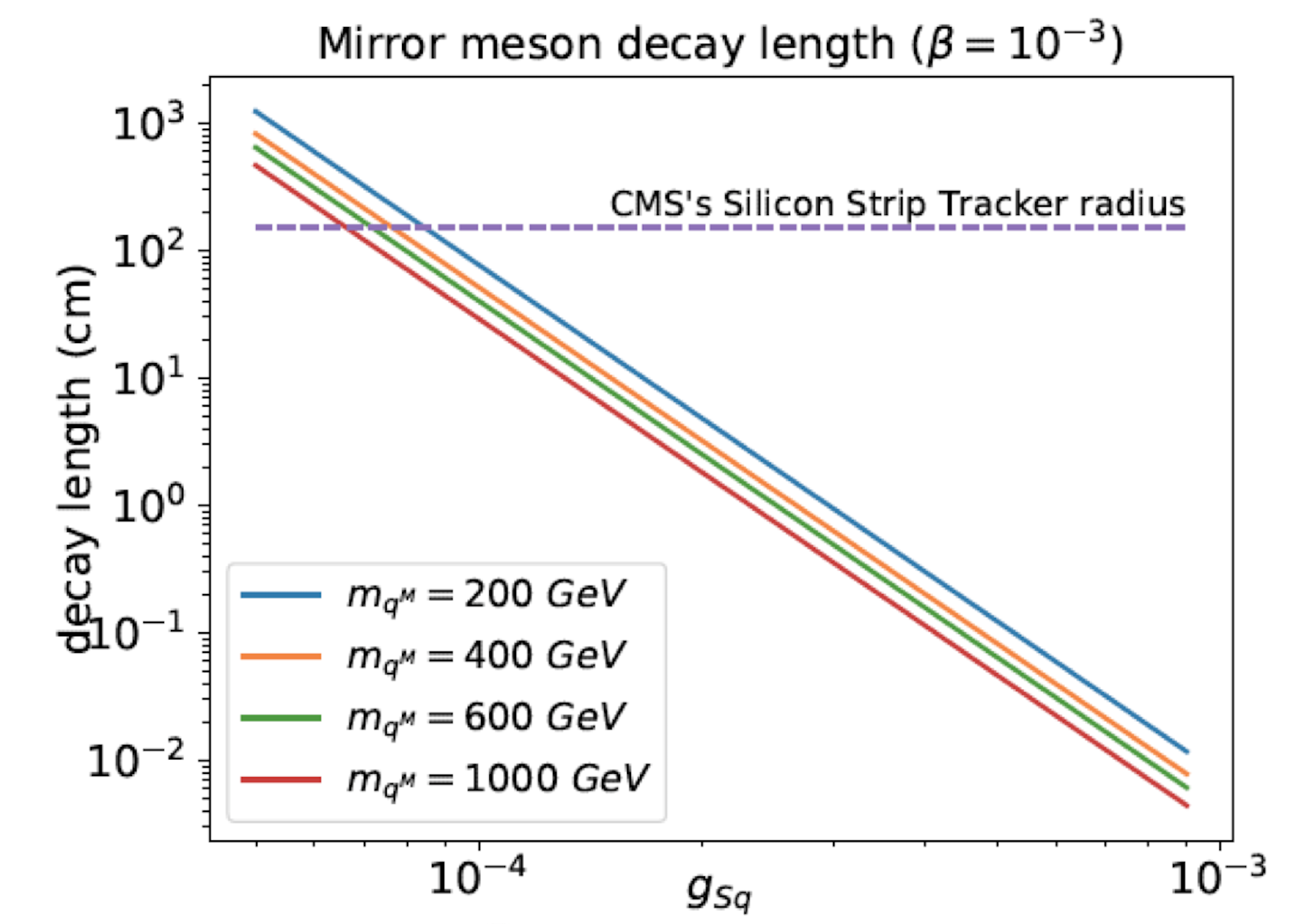}
 \caption{Figure shows the variation of the decay length (cm) with the generic coupling $g_{Sq}$ varying in the range of $10^{-3}$ to $10^{-4}$ for four different values of mirror quark masses ($M_{q^M} = 200, 400, 600$ and $1000$ GeV) with fixed $\beta$. Any macroscopic decay length of the mirror quarks above the broken black line (decay length 1mm) corresponds to the CMS's silicon strip tracker radius limit.}
\end{figure}
\subsection{{Mirror Lepton signatures}}
Let us now discuss the collider signatures of mirror leptons (charged mirror leptons, $e^{M\pm}$, as well as right handed mirror neutrinos, $\nu_R^M$ (for details refer to \cite{Chakdar:2016adj}). In this framework of {\em EW$\nu_R$} model, since the masses of mirror leptons being restricted to be in the ballpark of few hundred GeVs from the perturbativity of the Yukawa couplings, the pair production cross-section of the mirror leptons could be significant enough to probe or ruled out the {\em EW$\nu_R$} model at the ongoing/future runs of the LHC. As the mirror leptons have gauge coupling with photon, $W^\pm$ and $Z$-boson, therefore pair-productions of mirror leptons at the LHC take place through quark antiquark initiated processes with a $\gamma/W^\pm/Z$ in the s-channel. For example, the pair production of charged mirror leptons, $\bar e^M e^M$, (right handed mirror neutrinos, $\nu^M_R\nu^M_R$) proceeds via a photon or $Z$-boson exchange in the s-channel, whereas, $e^M \nu_R^M$ production takes place via $W^\pm$ exchange. After being produced, the mirror leptons decay into SM quarks, leptons, neutrinos and the singlet scalar, $\phi_S$. The final state neutrinos and $\phi_S$ remain elusive at the detector and thus, give rise to missing energy signature. Assuming $\nu_R^M$ being heavier than charged mirror lepton ($e^M$), there are two possible decay modes for the $\nu_R^M$. It can decay into a SM neutrino ($\nu_L$) and $\phi_S$, taking place via the relevant Yukawa interaction and hence, suppressed due to Yukawa coupling $g_{Sl}<10^{-4}$. $\nu_R^M$ dominantly decays into a $e^{M\pm}$ in association with a on/off shell (depending on the $e^M$--$\nu_R^M$ mass splitting)  $W^\mp$ which subsequent decays into a pair of jets or lepton-neutrino pair. The decay of $e^{M\pm}$ into a $W^\pm$ and $\nu_R^M$ is kinematically forbidden for $m_{e^M}<m_{\nu_R^M}$. Therefore, $e^{M\pm}$ decays into $e^\pm$ and $\phi_S$ with 100\% branching ratio.
In summary, the distinguished signals to look for are lepton-number violating signals at high energy: $q\bar{q} \rightarrow Z \rightarrow \nu_Ri+\nu_Ri \rightarrow e_{Lk} +e_{Ll} +W^+ +W^+ +\phi_S +\phi_S$: Like-sign dileptons $e_{Lk} +e_{Ll}$ + 2 jets (from 2 W) +missing energies (from $\Phi_S$) implies Lepton-number violating signals! The appearance of like-sign dileptons $(e^-e^-,\mu^-\mu^-,\tau^-\tau^-,e^-\mu^-,...)$ could occur at displaced vertices $> 1mm$ or even tens of centimeters depending on the size of $g_{Sl}$, which is an exciting feature of this framework.
\section{{Conclusions}}
As explained, the uniqueness of this framework lies in the fact that the gauge symmetry is the same as the SM, but it contains $\nu_R$ as well as mirror quarks and leptons at the EW scale. The model was invented to explain the tiny neutrino masses with EW scale $\nu_R$ masses, satisfies the EW precision data as well as all the constraints coming from the 125 GeV Higgs and MEG $\mu \rightarrow e \gamma$ constrains. It incorporates a rich scalar sector with additional mirror doublets, two Higgs triplets and singlet Higgses, which are under detailed study currently \cite{Chakdar et all}. Most excitingly, the $EW\nu_R$ model belongs to a class of models containing characteristic LLP signatures with low backgrounds and large displaced vertices (mm-cm) both in quark and lepton sectors. In the end, the framework presents a scenario where whether the Dirac or Majorana nature of neutrinos can be settled if like- sign di-lepton signals are discovered at displaced vertices at hadron colliders such as the LHC and presents a new theoretical direction in the post-125 GeV Higgs LHC.


\begin{thebibliography}{99}  
  \bibitem{Hung:2006ap}
P. Q Hung,
Phys. Lett. B \textbf{649}, 275-279 (2007)
doi:10.1016/j.physletb.2007.03.067
[arXiv:hep-ph/0612004 [hep-ph]].

\bibitem{Hung:2015nva}
P. Q Hung and T.~Le,
JHEP \textbf{09}, 001 (2015)
doi:10.1007/JHEP09(2015)001
[arXiv:1501.02538 [hep-ph]].

\bibitem{Hoang:2014pda}
V.~Hoang, P.~Q.~Hung and A.~S.~Kamat,
Nucl. Phys. B \textbf{896}, 611-656 (2015)
doi:10.1016/j.nuclphysb.2015.05.007
[arXiv:1412.0343 [hep-ph]].

\bibitem{Duong}
Dat Duong and P. Q. Hung, in preparation [arXiv:2006.xxxx[hep-ph]].

\bibitem{Chakdar:2015sra}
S.~Chakdar, K.~Ghosh, V.~Hoang, P. Q Hung and S.~Nandi,
Phys. Rev. D \textbf{93}, no.3, 035007 (2016)
doi:10.1103/PhysRevD.93.035007
[arXiv:1508.07318 [hep-ph]].

\bibitem{Chakdar:2016adj}
S.~Chakdar, K.~Ghosh, V.~Hoang, P. Q~Hung and S.~Nandi,
Phys. Rev. D \textbf{95}, no.1, 015014 (2017)
doi:10.1103/PhysRevD.95.015014
[arXiv:1606.08502 [hep-ph]].


\bibitem{Chakdar et all}
S. Chakdar, Dilip Ghosh, P.~Q.~Hung and, N. Khan
in preparation
[arXiv:2007.xxxx[hep-ph]].
  
 \end{thebibliography}
\end{document}